# Super-heavy fermion material as metallic refrigerant for adiabatic demagnetization cooling

Y. Tokiwa[1,2,3*], B. Piening[1], H. S. Jeevan[1], S. L. Bud'ko[4], P. C. Canfield[4], P. Gegenwart[1,3]

[1]I. Physikalisches Institut, Georg-August-Universität Göttingen, 37077 Göttingen, Germany.
[2]Department of Physics, Kyoto University, Kyoto 606-8502, Japan
[3]Experimental Physics VI, Center for Electronic Correlations and Magnetism, University of Augsburg, 86159 Augsburg, Germany.
[4]Ames Laboratory, US DOE and Department of Physics and Astronomy, Iowa State University, Ames, Iowa 50011, USA

*Corresponding author; email: yoshifumi.tokiwa@physik.uni-augsburg.de

## Abstract

Low-temperature refrigeration is of crucial importance in fundamental research of condensed matter physics, as the investigations of fascinating quantum phenomena, such as superconductivity, superfluidity and quantum criticality, often require refrigeration down to very low temperatures. Currently, cryogenic refrigerators with $^3$He gas are widely used for cooling below 1 Kelvin. However, usage of the gas is being increasingly difficult due to the current world-wide shortage. Therefore, it is important to consider alternative methods of refrigeration. Here, we show that a new type of refrigerant, super-heavy electron metal, $YbCo_2Zn_{20}$, can be used for adiabatic demagnetization refrigeration, which does not require $^3$He gas. A number of advantages includes much better metallic thermal conductivity compared to the conventional insulating refrigerants. We also demonstrate that the cooling performance is optimized in $Yb_{1-x}Sc_xCo_2Zn_{20}$ by partial Sc substitution with $x\sim0.19$. The substitution induces chemical pressure which drives the materials close to a zero-field quantum critical point. This leads to an additional enhancement of the magnetocaloric effect in low fields and low temperatures enabling final temperatures well below 100 mK. Such performance has up to now been restricted to insulators. Since nearly a century the same principle of using local magnetic moments has been applied for adiabatic demagnetization cooling. This study opens new possibilities of using itinerant magnetic moments for the cryogen-free refrigeration.

## Introduction

There have been various reports on the recent $^3$He crisis due to the increasing imbalance of demand and supply (*1-3*), which affects a variety of applications, including medical, military and scientific usages. The demand has been rapidly expanded due to increasing use from various applications, including neutron detectors in homeland security, while the supply is limited since the gas is produced only through tritium decay in nuclear weapon stockpile and nuclear reactors (*1,2*). The increasing demand is becoming unsustainable, leading to a steep rise in the price by a factor of more than 10 from 2007 to 2009 (*3*). This crisis also impacts the field of condensed matter physics, since $^3$He cryogenic refrigerators are most commonly used for cooling below 1 K. Therefore, finding alternative refrigeration techniques is an urgent issue in the field. One of the possible candidates for replacing such cryogenic refrigerators is adiabatic demagnetization refrigeration (ADR) (*4*).

Current ADR for sub-K cooling uses paramagnetic insulators as refrigerants, so-called paramagnetic salts with local magnetic moments (*4-5*). At zero field the moments are randomly oriented, whereas at high fields the moments are aligned with reduced magnetic entropy (Fig.1A). When the magnetic field is decreased from high field to zero, disordered magnetic moments with increased entropy absorb heat from the lattice, leading to demagnetization cooling. The measure of effectiveness in cooling is the magnetocaloric effect (MCE, $\partial T/\partial H|_S$), which quantifies the change of temperature caused by a change of the magnetic field in adiabatic conditions. Paramagnetic salts are widely used because of their large MCE in the temperature range below a few K down to a few tens of mK. However, their poor thermal conductance due to their insulating nature prevents an effective low-$T$ heat transport. The same holds true for low-dimensional spin chains, tuned towards a quantum critical point (see below) (*6*). Recently, ADR to ~0.2 K using the metallic compound $YbPt_2Sn$ has been demonstrated (*7*). In this compound the magnetic moments of Yb-ions, which are responsible for cooling, are fully localized and exhibit a very small mutual interaction. Thus, despite being metallic, it relies on the same principle of operation as paramagnetic salts. Furthermore, the co-existing conduction electrons lead to a small but finite RKKY interaction, which causes magnetic ordering below 250 mK, preventing the cooling to lower temperatures. Since the proposal by P. Debye nearly a century ago (*8*), the same principle of using local magnetic moments for ADR has been applied. Below, we show a completely new approach to ADR, which is based upon using itinerant magnetic moments in a super-heavy electron (HE) metal. Our example demonstrates cooling well below 0.1 K.

A HE state is formed at temperatures below the Kondo temperature $T_K$ through an exchange interaction between *f*- and conduction electrons (*9-11*). The HE state is easily destroyed by external parameters, such as magnetic field and pressure, owing to its low energy scale $k_B T_K$. Because the electronic entropy of a metal is proportional to the electronic density of states at the Fermi energy, $D(E_F)$, that is proportional to the effective electron mass $m^*$, the HE state has very large entropy (at zero field). This large entropy is suppressed as the HE state is destroyed by applying a magnetic field of order $\sim\mu_0 H \sim k_B T_K/g\mu_B$ (Fig.1A). Thus, with decreasing field from high values to zero a very rapid increase of the entropy is expected. The latter will absorb heat from the lattice system, implying that HE materials can be used for ADR.

HE systems with effective doublet ground state contain a magnetic entropy of $R\ln(2)$ at $T_K$. In order to optimally use all available magnetic entropy for ADR, the best parameters are: the initial temperature $T_i \sim T_K$, initial field $\mu_0 H_i \sim k_B T_K/g\mu_B = \mu_0 H_K$ and final field $H_f = 0$. Here, $g$ is the g-factor of conduction electrons. For prototypical HE systems, such as $YbRh_2Si_2$, both $T_K \sim 25$ K (*12*) and $\mu_0 H_K \sim 10$T (*13*) are much too large for ADR applications which typically require $T_i$ =1-2 K, $\mu_0 H_i \leq 8$ T. On the other hand, for Kondo lattice metals with low $T_K \sim 1$ K, typically, the exchange coupling between the moments leads to magnetic ordering, before the full moments are Kondo screened upon cooling. For Ce-based HE systems, the RKKY interaction is typically of order 10 K, while for Yb-based Kondo metals, it could be weaker. In order to retain a paramagnetic Kondo screened state for an ADR refrigerant material with a low Kondo temperature of only 1 K, a very weak RKKY interaction energy is required. A careful scan of available results in literature has revealed one most suitable system, $YbCo_2Zn_{20}$ (*14*) (Fig.1B), which seems to fulfill these severe requirements.

Among all known paramagnetic HE metals, it has one of the lowest Kondo temperatures, $T_K$ =1.5 K (*14*) (HE metals with such small or even smaller $T_K$ display usually a long-range ordered ground state). Reflecting the very low $T_K$, the Kondo coherent state is formed only at very low temperatures, as evidenced by the maximum at 2.5 K in electrical resistivity shown in Fig.1C. We ascribe the extraordinarily small $T_K$ and $T_{RKKY}$ to its crystal structure, which is the cubic

CeCr$_2$Al$_{20}$ type shown in Fig.1B, consisting of Zn cages surrounding the Yb atoms (*14*). These cages lead to a very weak hybridization between the Yb 4*f* with the Co 3*d* electrons. This effectively reduces the magnetic exchange interaction *J*, which enters both $T_K$ and $T_{RKKY}$. The paramagnetic Fermi liquid (FL) ground state is evidenced by specific heat measurements, which show rapid increases of *C/T* with decreasing temperature below $T_K$ =1.5 K and saturation at ~8 J/mol K$^2$ (cf. Fig. 2), i.e. one of the largest values among heavy fermion materials (10,11,13). This super HE state is very effectively suppressed by magnetic field, as indicated by rapid decrease of the $T^2$-coefficient in electrical resistivity (*15*). These properties set the system as a most promising candidate for a super-HE refrigerant. It should be also noted that YbCo$_2$Zn$_{20}$ exhibits a field-induced quadrupole ordering for the field only along the [111] direction (*16*). Because this ordering prevents the rapid decrease of the Sommerfeld coefficient, effective cooling is achieved only when the field is applied away from the [111] direction.

Since the magnetic Grüneisen ratio $\Gamma_H = T^{-1}(\partial T/\partial H)_S$, which equals the adiabatic magnetocaloric effect, diverges at any field-sensitive QCP (*17*), the usage of quantum critical materials as ADR refrigerants was proposed previously (*6*). It has been reported previously by electrical resistivity under pressure, that YbCo$_2$Zn$_{20}$ can be tuned to a QCP around 1 GPa, beyond which long-range antiferromagnetic ordering is found (*18*). We therefore expect that the ADR performance of YbCo$_2$Zn$_{20}$ could be further enhanced when tuning this material towards its QCP. However, for any practical application, the need of applying a hydrostatic pressure leads to severe complications. Therefore, we have been investigating, if chemical pressure can be used to drive YbCo$_2$Zn$_{20}$ towards quantum criticality. Indeed we found, that partial substitution of Yb by the smaller Sc leads to an effective chemical pressure. Below, we investigate the possibility of using the super HE system YbCo$_2$Zn$_{20}$ for ADR and study also the effect of quantum criticality, induced by chemical pressure, on the cooling performance.

**Results**

**Characterization**

Using flux growth (see methods), single crystals of Yb$_{1-x}$Sc$_x$Co$_2$Zn$_{20}$ with *x*=0, 0.13 and 0.19 were grown. The evolution of the lattice constant follows the Vegard's law, indicating successful substitution of Sc and the significant lattice contraction with substitution displays a clear chemical pressure effect (see supplemental material). The temperature dependence of the electrical resistivity of these materials is shown in Fig. 1C. The maximum around 2 K for *x* =0 indicates the crossover to the coherent Kondo lattice state at low temperatures. In the Sc substituted material with *x* =0.13 the resistivity maximum is shifted to lower temperature, suggesting a suppression of $T_K$ by the effect of chemical pressure. Furthermore, the residual resistivity ρ(*T*→0) is substantially enhanced due to the structural disorder caused by chemical substitution. The resistivity maximum is completely suppressed by further Sc substitution of *x*=0.19. Furthermore, the residual resistivity ρ(*T*→0) is substantially enhanced due to the structural disorder caused by chemical substitution. The resistivity maximum is completely suppressed by further Sc substitution of *x*=0.19. Here, we note that such enhancement of electrical resistivity would benefit in ADR applications, because it reduces eddy current heating in the demagnetization process. The un-substituted material may not be used for application because of the very high purity, leading to the observation of the de Haas-van Alphen effect (*16*). We show here that thermal conductivity of Sc-substituted materials is still higher than that of paramagnetic salts. By using the Wiedemann-Franz law with the resistivity value at the maximum ~70 μΩcm, we estimate the smallest electronic thermal conductivity at 50 mK to be $2\times10^{-5}$ W/Kcm, which is larger than a typical value of paramagnetic salts $7\times10^{-6}$ W/Kcm (*4,19*).

Figure 2 shows the electronic specific heat divided by temperature, $C_{el}/T$, of $Yb_{1-x}Sc_xCo_2Zn_{20}$. The data for $x=0$ increase logarithmically as temperature is decreased and reach a very large constant value of ~7 J/mol K$^2$ at 0.2 K, indicating the formation of a super HE state. Chemical pressure by Sc-substitution suppresses FL behavior. Indeed the heat capacity coefficient for $x=0.13$ and 0.19 display a further approximately logarithmic increase upon cooling below 0.2 K. Such behavior is a signature of quantum criticality, as found in many other HE materials (*20*). At the lowest measured temperature of 0.05 K, $C_{el}/T$ reaches a value as large as 8.5 J/mol K$^2$. Similar to unsubstituted $YbCo_2Zn_{20}$ (*16*), $C_{el}/T$ for $x=0.19$ is reduced by almost two orders of magnitude by the application of a magnetic field of 8T. Consequently, the large low-temperature electronic entropy found in zero field is almost completely shifted up to high temperatures by magnetic field (inset in Fig.2). This large entropy difference enables efficient ADR from 1.5 K at 8 T to 0.075 K at 0T. We note that the difference of entropy is 90% of $R\ln(2)$ in this cooling process, indicating cooling efficiency cannot be much better in any material, as long as the magnetic entropy of doublet ground states is used. In contrast to paramagnetic salts for which the specific heat follows the high-temperature tail of a Schottky anomaly $C/T \sim 1/T^3$ and thus rapidly decreases upon warming, $Yb_{1-x}Sc_xCo_2Zn_{20}$ with the milder $\ln(T)$-dependence displays huge $C/T$ values up to high temperature of ~2 K. As pointed out previously in the context of insulating quantum critical magnets, this large value of specific heat at high temperatures leads to a significantly longer "hold" time compared to paramagnetic salts, which could be of advantage for applications (*7*).

**Quantum criticality of Sc-substituted material**

In order to compare the performance of the three different samples for ADR, we studied in detail the adiabatic magnetocaloric effect, which equals the magnetic Grüneisen parameter, $\Gamma_H = T^{-1}(\partial T/\partial H)_S$, utilizing the alternating field method (*21*). For quantum critical materials, this quantity is expected to display divergent behavior, which has been confirmed for several quantum critical materials, recently (*17,22-24*). We stress that the nature of a QCP is characterized by $\Gamma_H$ much more clearly compared to $C/T$. This is, because $\Gamma_H$ is dominated by the quantum critical contribution with strongest $T$-dependence, while $C/T$, e.g. for antiferromagnetic quantum criticality in three dimensions is dominated by the non-critical FL contribution, which in our case is extremely large ($\gamma \sim 7$ J/mol K$^2$). Electrical resistivity is also not a good probe of quantum criticality in this study, since the resistivity of the Sc-substituted materials has a large residual contribution as shown in Fig.1.

Figure 3 shows temperature and magnetic field dependencies of $\Gamma_H$ for $x=0$, 0.13 and 0.19. $\Gamma_H(T)$ for $x=0$ increases strongly with decreasing temperature and passes through a maximum around 0.5 K. Since $\Gamma_H = -(dM/dT)/C$ (*M*: magnetization), the non-monotonic behavior of $\Gamma_H(T)$ for $x=0$ is probably related to the o bserved maximum in magnetic susceptibility at 0.3 K (*15*). As shown in Fig. 3B, the field dependence of $\Gamma_H$ for $x=0$ at $T=80$ mK changes sign at 0.5T. Such sign change is a typical signature of metamagnetic behavior and consistent with the previously reported metamagnetism at 0.5 T (*15*), which is likely ascribed to itinerant-moment magnetism, since 4*f*-electrons are itinerant in this system. Upon partial substitution of Yb by Sc, the non-monotonic behaviors in temperature and field dependencies weaken for $x=0.13$ and completely disappear for $x=0.19$. $\Gamma_H$

($T$) at $\mu_0H$ =0.1 T for $x$=0.19 diverges with $T^{-1\pm0.05}$ below ~0.6K and $\Gamma_H(H)$ above $\mu_0H$ =1 T follows $0.52/\mu_0H$. The $H^1$ divergence instead of $(H-H_{QCP})^{-1}$ indicates that the system is tuned exactly to the QCP with $H_{QCP}$ =0. The $T$-exponent and the obtained $1/H$ prefactor $0.52\pm0.03$ agree very well with the theoretical prediction for a zero-field QCP of an AFM spin-density-wave (SDW), which yields $\Gamma_H(T,H) \sim H\, T^1$ in the quantum critical (QC) regime and $\Gamma_H(T,H)$ =$0.5/\mu_0H$ in FL regime (*17,24*). This theoretical prediction indicates the appearance of a maximum in $\Gamma_H(H)$, corresponding to a cross-over field between QC and FL regimes, because $\Gamma_H(H)\sim H$ and $\Gamma_H(H)\sim 1/H$ in the former and the latter regimes, respectively. The positions of such maxima is shown by the dotted lines in Fig.4. For $x$=0.19 the deviation of $\Gamma_H(H)$ below $\mu_0H$ =1T from the expected $1/H$ divergence is caused by the gradual cross-over from FL to QC regimes with decreasing field. It is intriguing that the system follows exactly the theoretical prediction of a SDW QCP, even though it contains significant dilution of magnetic Yb sites. Another likely possibility for causing quantum critical behavior is a quantum Griffiths phase, which is expected in systems with large amount of disorder. However, the observed critical behavior is incompatible with Griffiths phase, since theory predicts for this case only a $\log(T)$ dependence in the Grüeneisen parameter, which is much weaker than $T^{-1}$ (*25*).

Detailed measurements of $\Gamma_H$ are summarized in the color-coded plot, shown in Fig. 4. There appears a small area of negative $\Gamma_H$ for $x$=0 due to the sign change in its field dependence. This area of sign change is suppressed as Sc is substituted and monotonic field/temperature dependence is found for $x$=0.19. The strongest enhancement towards the origin reflects the QCP at zero field for this composition. It is evident, that the $x$=0.19 sample is best suited for ADR towards zero magnetic field, because the region of highest $\Gamma_H$ extends towards the origin for this material. We also note, that further growth of $\Gamma_H(T)$ upon cooling to even lower temperatures is expected within the quantum critical regime at $H$=0.

**Adiabatic demagnetization cooling by super-heavy electrons**

Using these data, we determined the cooling curves under adiabatic conditions (Fig. 5). All the curves display significant cooling effects. The curve for $x$=0.19 shows a monotonic demagnetization cooling, whereas it is non-monotonous for the unsubstituted material at low temperatures with a minimum around 0.6T, which is caused by the metamagnetism (*26*). When starting from $T_s$=2.5 K and $\mu_0H_s$=8 T, the cooling effect for $x$=0 is slightly better than that for $x$=0.13 and 0.19 with a lower final temperature at zero field ($T_f$). Starting from $T_i$ =1 K, $T_f$ becomes slightly lower with steeper slope at low temperature for x=0.13 and 0.19 than x=0. As shown in the inset of Fig. 2, steeper slope of entropy, which equals $C/T$, leads to a lower final temperature. Since $C/T$ is larger for $x$=0.13 and 0.19 than $x$=0 below $T$~100 mK, better cooling due to the QCP is realized at very low temperatures. We point out that $C/T$ may diverge towards zero temperature at the QCP. Thus, cooling to arbitrary low temperatures may be possible, due to the significant $dS/dT$= $C/T$ even in the limit of zero temperature. For $x$=0.19, in fact, the final temperature reaches the lower limit of our experimental setup, 40 mK, when we set the initial temperature to a low value $T_i$ =1 K.

**Discussion**
In this study we show that the HE metallic systems $Yb_{1-x}Sc_xCo_2Zn_{20}$ can be used as new type of refrigerant for ADR with itinerant magnetic moments. The steep increase of entropy with temperature at very low temperature due to the diverging $C/T$ in the quantum critical Sc-substituted material (Fig. 2) enables cooling down to temperatures even lower than the lowest temperature of our experimental setup, 40 mK. These results, along with the usage of most of the available magnetic entropy $R\ln(2)$ (inset of Fig. 2), indicate that the quantum critical Sc-substituted $YbCo_2Zn_{20}$ is an ideal metallic refrigerant. Heavy electron metallic refrigerants have major advantages compared to insulating paramagnetic salts, such as an additional electronic thermal conductivity, stability in the air and easy machining since they are much less brittle compared to salts. In the conventional ADR systems, large amounts of Cu or Ag wires are spread uniformly in the paramagnetic salt pill to improve thermal conductance (*4*). One possible immediate application may be replacing the metallic wires with $YbCo_2Zn_{20}$, which would improve the cooling performance strongly. The HE refrigerants, thus, provide realistic possibilities of application and even replacement for conventional paramagnetic salts.

**Materials and Methods**

Single crystals of $Yb_{1-x}Sc_xCo_2Zn_{20}$ were grown by the self-flux method (*14,27*). The compositions of obtained crystals were determined by energy dispersive X-ray analysis.

Electrical resistivity was measured using the standard four-probe technique.

Single crystals of $Yb_{1-x}Sc_xCo_2Zn_{20}$ with dimension of ~ 2 x 2 x 0.5 mm$^3$ were used for specific heat and magnetocaloric effect (MCE) experiments.

The specific heat was measured by the standard quasi-adiabatic heat pulse method, using a dilution refrigerator.

The MCE was measured by utilizing an alternating field technique adapted to a dilution refrigerator (*21*). We varied frequency (*f*) for the alternating field, depending on temperature. The sample is in a quasi-adiabatic condition with a weak heat link to the bath, causing *T*-dependent relaxation time $\tau$, which increases with decreasing *T*. To ensure the effective adiabatic condition for MCE measurements, it is necessary to maintain a condition, $f \gg 1/\tau$. We also point out that the relaxation time between the sample and thermometer ($\tau'$) becomes longer with decreasing *T*, leading to the other necessary condition, $f \ll 1/\tau'$. Thus one needs to maintain $\tau' \gg 1/f \gg \tau$. This is easily checked by measuring the *f* dependence of the MCE signal. If the condition is maintained, the signal is *f*-independent. We obtained *f*-independent MCE signal for each measurement. Typically, we use 0.1 Hz at high temperatures above 1K and 0.02 Hz at low temperatures below 0.2 K. This ensures perfect adiabatic conditions. Our data of the MCE thus equal the magnetic Grüneisen parameter. We also changed the amplitude of the alternating field. Since the magnetic Grüneisen parameter becomes very sensitive to the field at low temperatures, we decreased the amplitude with decreasing temperature. We decrease it from 0.02 T at temperatures above 1K to 0.004 T below 0.2 K.

**Supplementary Materials**
Fig. S1. Sc-substitution dependence of lattice constant *a*

**References and Notes**
1. D. Shea, D. Morgan, The Helium-3 shortage: supply, demand, and options for congress. *Congressional Research Service* R41419 (2010).
2. R. T. Kouzes, J. H. Ely, Status summary of $^3$He and neutron detection alternatives for homeland security. (2010). Available at http://www.pnl.gov/main/publications/external/technical_reports/PNNL-19360.pdf.
3. A. Cho, Helium-3 shortage could put freeze on low-temperature research. *Science* **326**, 778–779

**Acknowledgments**

Financial support for this work was provided by the German Science Foundation and the Grants-in-Aid for Scientific Research (No. 15K13521) from the Japan Society for the Promotion of Science (JSPS). Part of this research was performed by P. C. Canfield and S. L. Bud'ko at the Ames Laboratory and supported by the U.S. Department of Energy, Office of Basic Energy Science, Division of Materials Sciences and Engineering. Ames Laboratory is operated for the U.S. Department of Energy by Iowa State University under Contract No. DE-AC02-07CH11358. All data needed to evaluate the conclusions in the paper are present in the paper and/or the Supplementary Materials. Additional data related to this paper may be requested from the authors.

**Author contributions:**

P.G. conceived the project. Y.T. and P.G. planed and designed the experiments. Measurements were performed by Y.T and B.P. The samples were synthesized and characterized by B.P., H.S.J., S.L.B. and P.C.C. Y.T. and P.G. discussed the results and prepared the manuscript.

**Competing interests:** The authors declare that they have no competing interests.

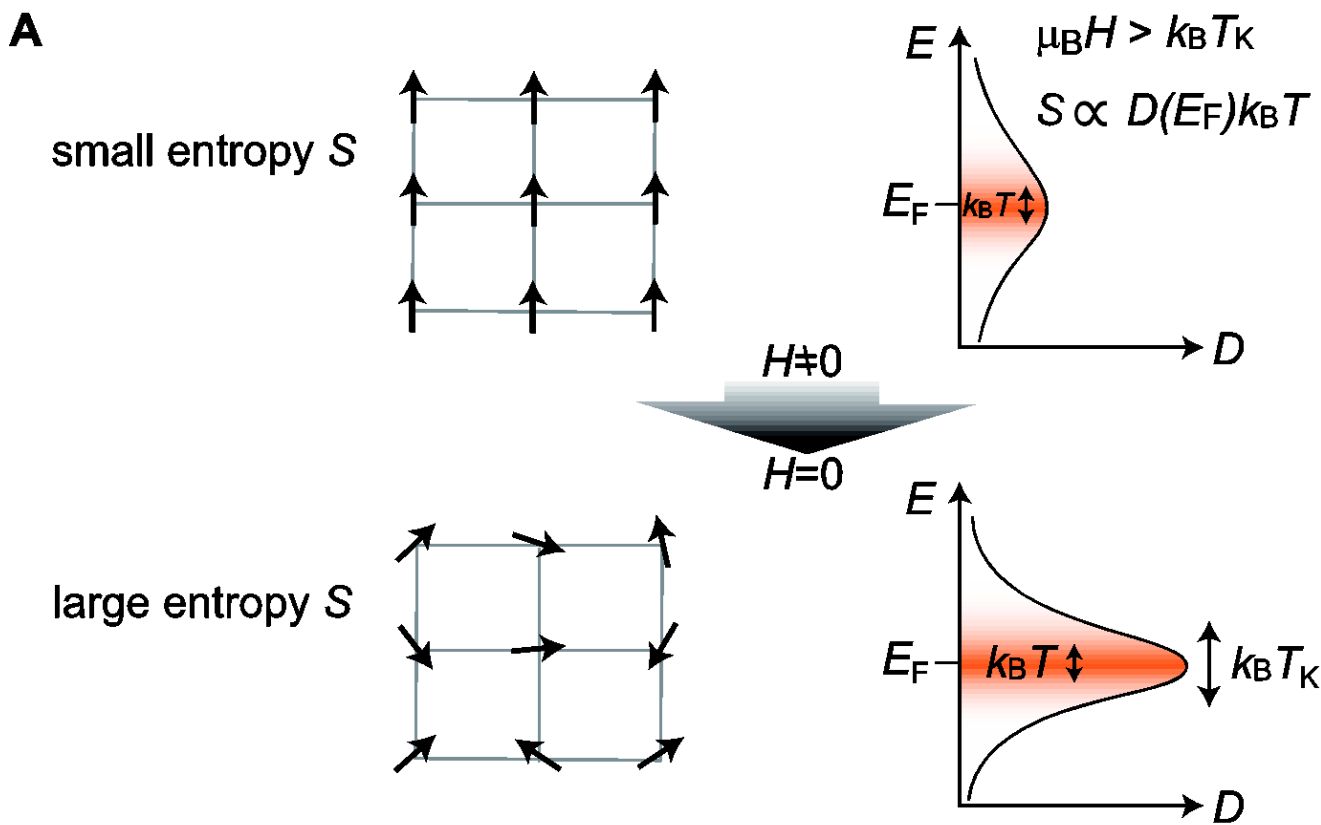
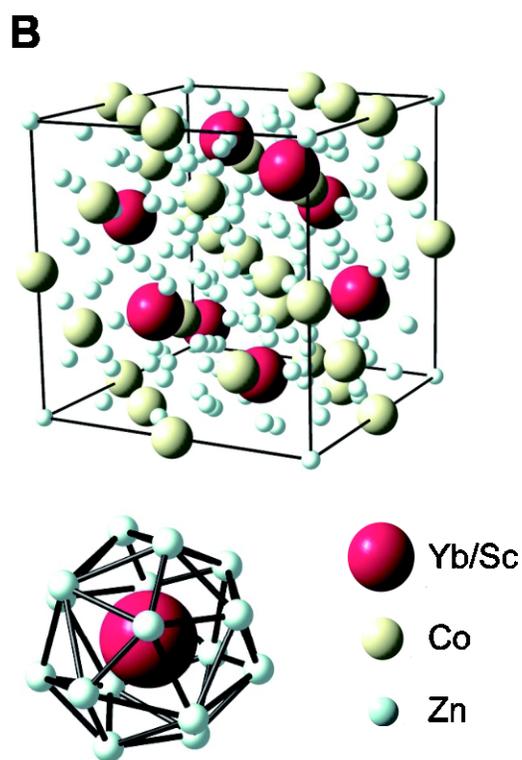
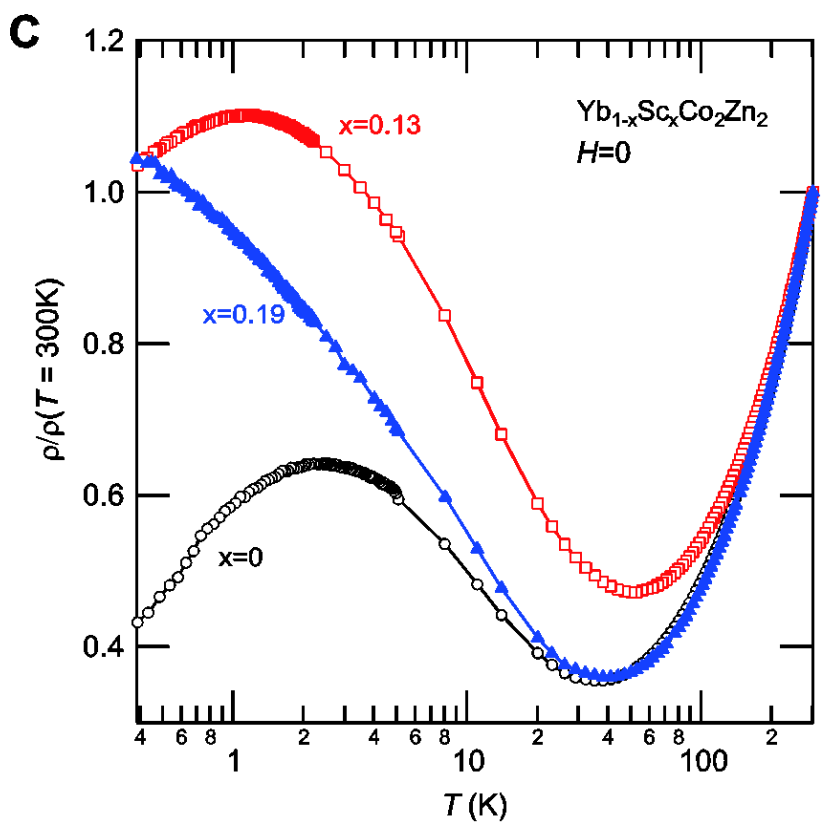

**Fig. 1. Adiabatic demagnetization cooling by a heavy electron refrigerant.** (**A**), Comparison of the demagnetization processes for conventional localized-moment (left) and heavy electron itinerant-moment (right) refrigerants. For the latter (right figures) the entropy $S$ is proportional to the number of thermally excited holes and electrons, $D(E_F)k_BT$, which corresponds roughly to the area depicted by orange color. Here, $D$ is the density of states. These materials at zero field have a very large and sharp peak with a width of ~ $k_BT_K$ in the density of states near $E_F$. The density of states is being strongly suppressed by the application of fields exceeding $\mu_0H = k_BT_K/g\mu_B$ (right upper) (*14,28-30*), which could be used for adiabatic demagnetization cooling. Note, that these are schematic sketches only and the true $D(E)$ in particular for $H \geq 0$ will display more fine structure. (**B**), Crystal lattice structure of the super-heavy electron refrigerant $YbCo_2Zn_{20}$ and the cage structure of Zn surrounding Yb. (**C**), Temperature dependence of electrical resistivity at zero field of $Yb_{1-x}Sc_xCo_2Zn_{20}$ with *x*=0, 0.13 and 0.19.

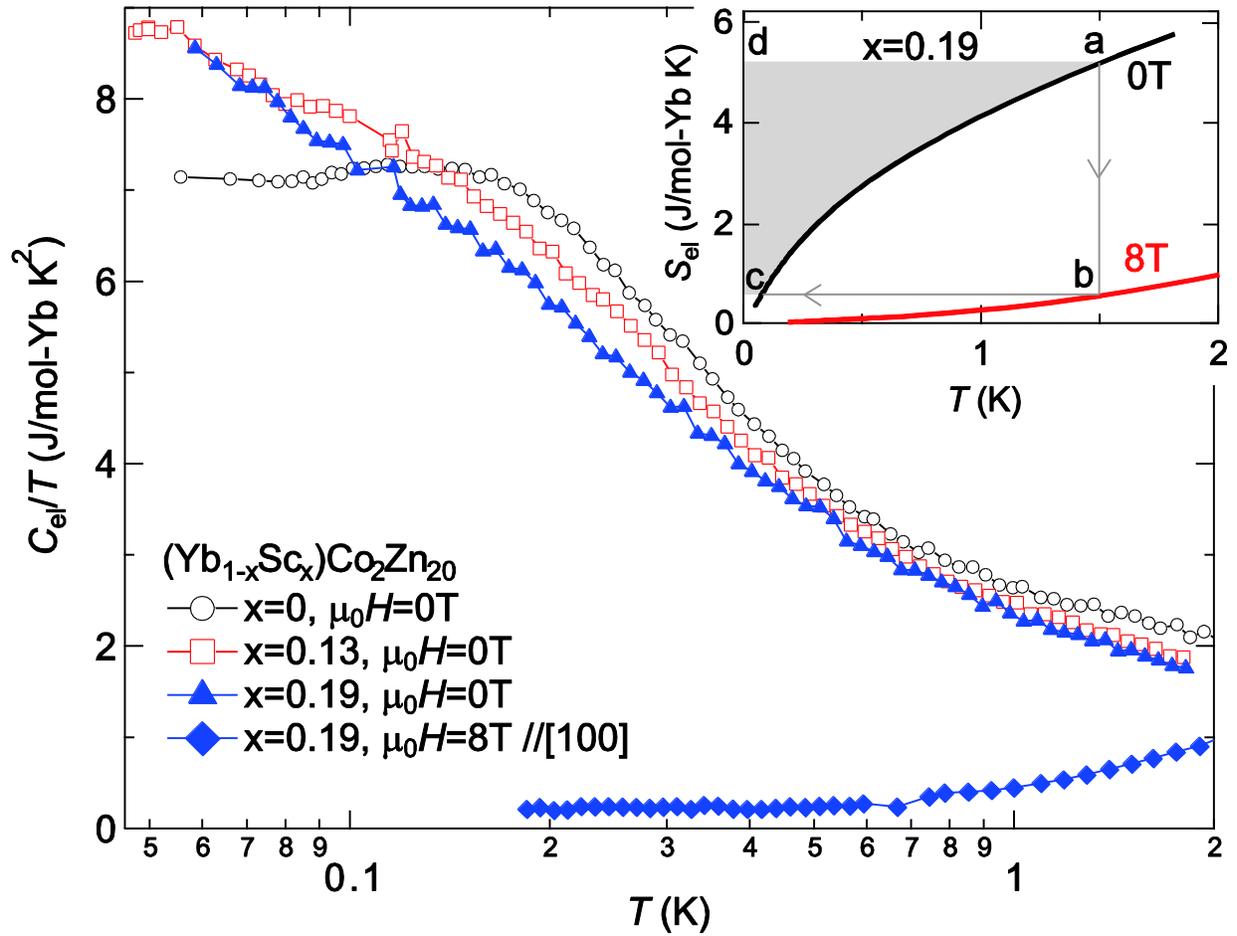

**Fig. 2. Formation of super-heavy electron state in $YbCo_2Zn_{20}$ and diverging effective mass in 19% Sc-substituted material.** Electronic specific heat divided by temperature $C_{el}/T$ for $H//[100]$ plotted against temperature. Nuclear specific heat was subtracted. $C_{el}/T$ at zero magnetic field for $x=0$, 0.13 and 0.19 plotted with black open circles, red open squares and blue solid triangles, respectively. Data for $x=0.19$ at a magnetic field of 8T along the [100] direction are indicated by blue solid diamonds. Inset shows the calculated electronic entropy for $x=0.19$ at zero field and 8 T under the assumption of constant $C_{el}/T$ below the lowest measured temperatures. It is noted that this assumption leads to a slight underestimation of entropy at zero field. The grey lines with arrows indicate the demagnetization cooling process, starting from $\mu_0 H=8$ T and $T=1.5$ K. The final temperature $T_f$ is 0.075 K. The grey area equals the amount of the heat $\Delta Q_c=2.2$ J/mole, which the material absorbs in the cooling process, while the area of rectangle a-b-c-d equals the heat $\Delta Q_m =5.6$ J/mole which is transferred from the material to heat bath. These yield a high efficiency factor $\Delta Q_m /\Delta Q_c$ of 40 %.

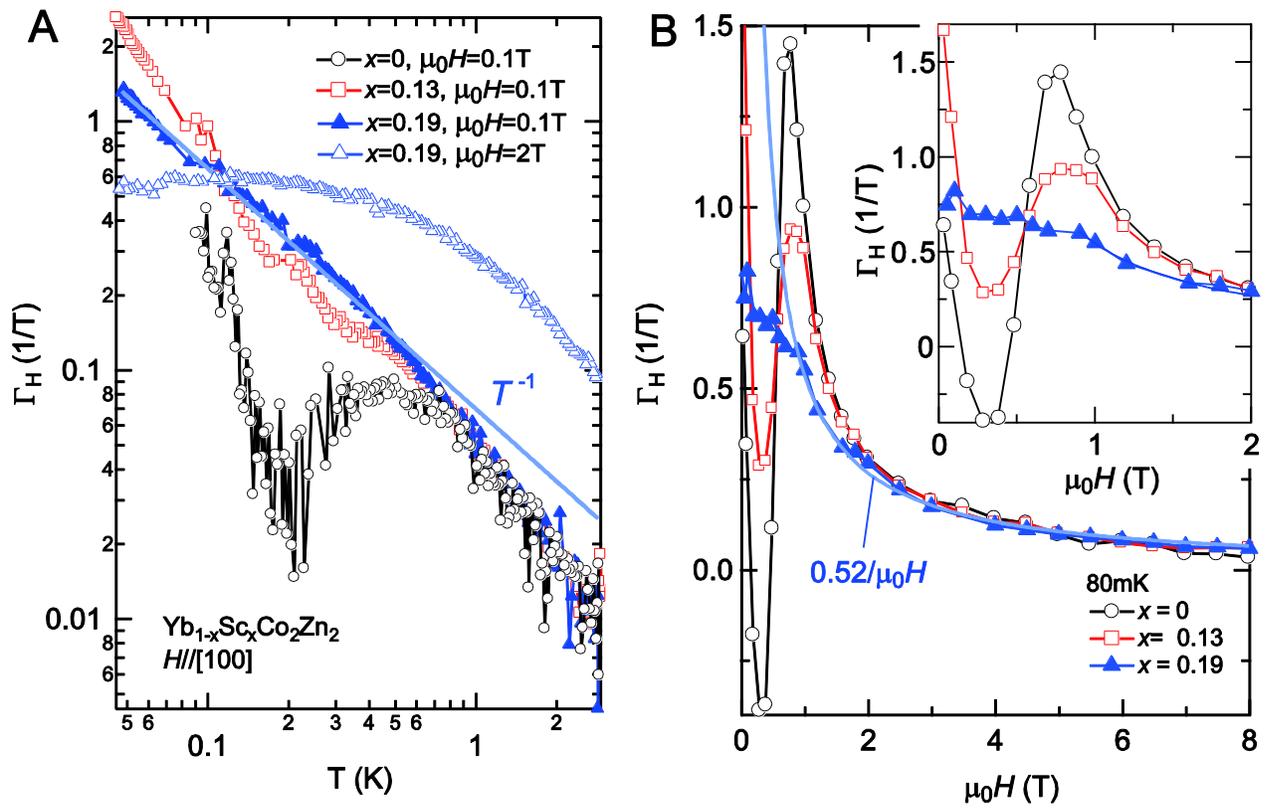

**Fig. 3. Tuning to QCP by Sc-doping in $Yb_{1-x}Sc_xCo_2Zn_{20}$.** (**A**), Magnetic Grüneisen ratio $\Gamma_H$ of $Yb_{1-x}Sc_xCo_2Zn_{20}$ with $x$=0, 0.13 and 0.19 as a function of temperature for $H//[100]$. The divergence with a power law ~ $T^{-1}$ for $x$=0.19 is indicated by the blue solid line. (**B**), $\Gamma_H$ at $T$=80 mK as a function of magnetic field. The solid blue line is a fit to the data for $x$=0.19 from $\mu_0 H$ =1 T to 8 T. The Inset is a blow-up of the main figure for the low-field region.

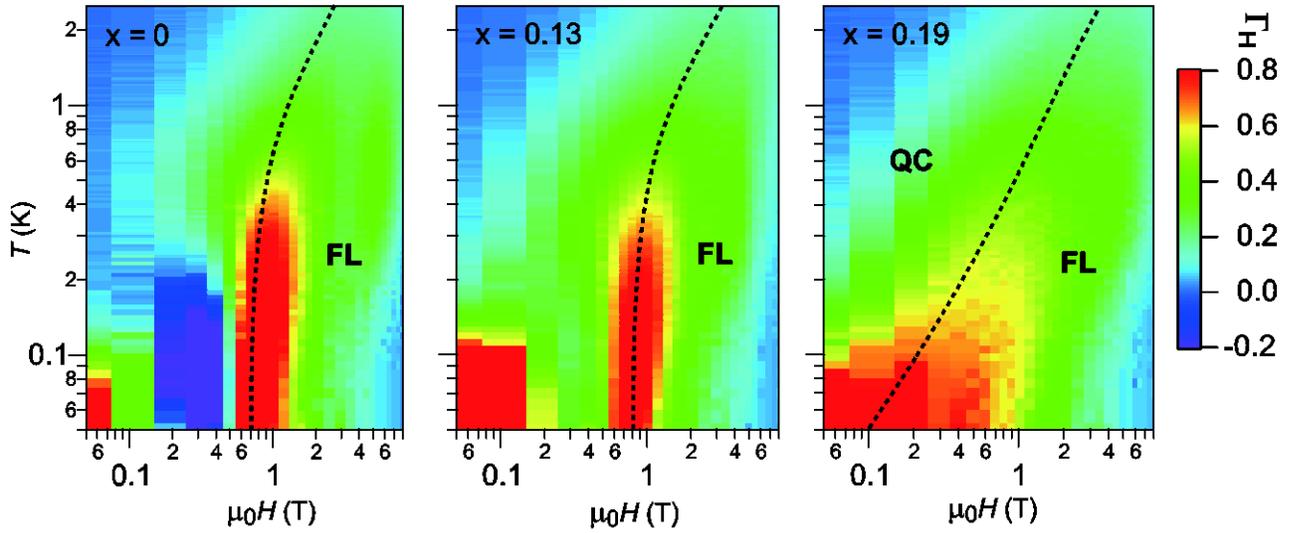

**Fig. 4. Visualization of tuning to the QCP by Sc-substitution.** Color-coded contour plot of the magnetic Grüneisen parameter $\Gamma_H$ of $Yb_{1-x}Sc_xCo_2Zn_{20}$ in $H$-$T$ phase space. Magnetic field has been applied parallel to the [100] direction. Dotted lines indicate maxima positions in the field dependence of $\Gamma_H(H)$. These lines correspond to the cross-over field to the Fermi liquid (FL) regime at high fields (24). (see the main text for explanation.) For $x$=0 and 0.13 the systems at low fields are influenced by quantum fluctuations of metamagnetism around 0.5T (16), causing a finite-field extrapolation of the maximum position in $\Gamma_H(H)$ for $T\rightarrow 0$. For the critical concentration $x_c$=0.19, the line is extrapolated to zero, reflecting a zero-field quantum critical (QC) point. $\Gamma_H$ obeys the expected QC behavior of spin-density-wave instability, namely $\Gamma_H(T)\sim 1/T$ in the QC regime and $\Gamma_H(H)\sim 1/H$ in the FL regime (18,24).

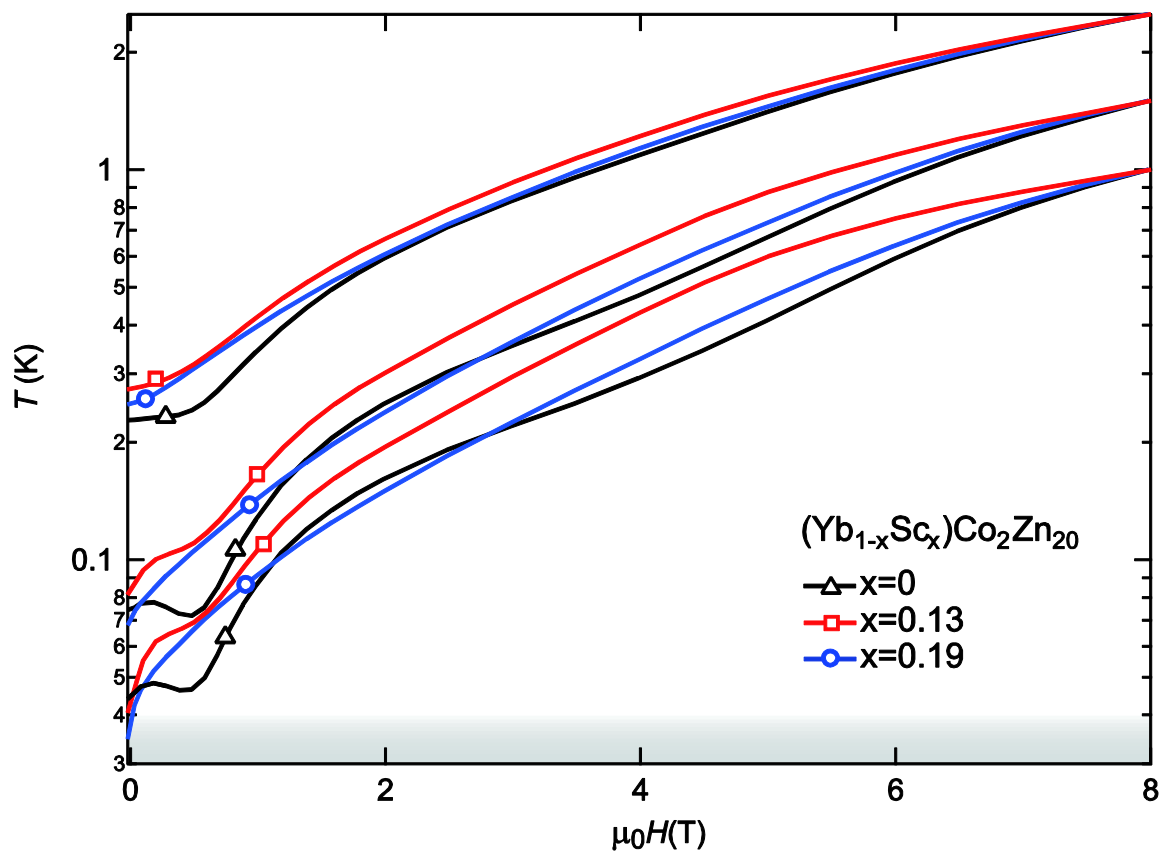

**Fig. 5. Adiabatic demagnetization refrigeration of $Yb_{1-x}Sc_xCo_2Zn_{20}$.** Solid black, red and blue curves represent the cooling curves for $x$=0, 0.13 and 0.19, respectively. The curves are obtained by integrating the magnetocaloric effect ($\partial T/\partial H|_S$) from 8 T to zero field. Magnetic field is applied parallel to the [100] direction. The points at low temperature below the lower limit of the thermometer calibration (40 mK) are obtained by extrapolating the calibration data. The out-of-calibration range is shaded by gray color.

# Supplementary Material

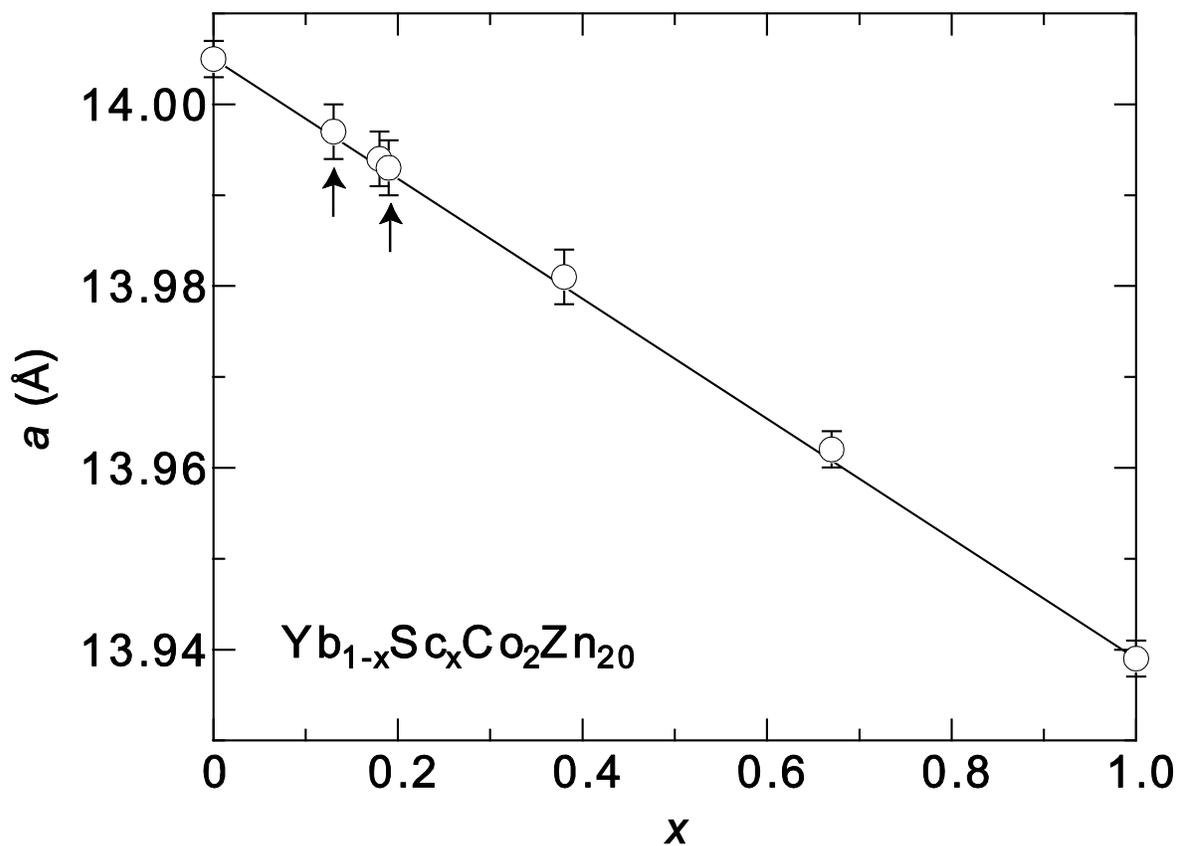

**Fig. S1. Sc-substitution dependence of lattice constant *a*.** The solid line indicates the volume compression according to Vegard's law. Samples with $x=0.13$ and $0.19$, indicated by arrows, have been used for this study.